\documentclass[11pt,a4paper]{llncs}
\usepackage{fullpage,graphicx}

\newcommand{\Oh}[1]
    {\ensuremath{\mathcal{O}\!\left( {#1} \right)}}
\newcommand{\SA}
    {\ensuremath{\mathrm{SA}}}
\newcommand{\SSA}
    {\ensuremath{\mathrm{SSA}}}
\newcommand{\prank}
    {\ensuremath{\mathrm{p\_rank}}}
\newcommand{\select}
    {\ensuremath{\mathrm{select}}}

\begin{document}

\title{Compressed Spaced Suffix Arrays}
\author{Travis Gagie\inst{1} \and
Giovanni Manzini\inst{2} \and
Daniel Valenzuela\inst{1}}
\institute{University of Helsinki, Finland \and
University of Eastern Piedmont, Italy}
\maketitle

\begin{abstract}
Spaced seeds are important tools for similarity search in bioinformatics, and using several seeds together often significantly improves their performance.  With existing approaches, however, for each seed we keep a separate linear-size data structure, either a hash table or a spaced suffix array (SSA).  In this paper we show how to compress SSAs relative to normal suffix arrays (SAs) and still support fast random access to them.  We first prove a theoretical upper bound on the space needed to store an SSA when we already have the SA.  We then present experiments indicating that our approach works even better in practice.
\end{abstract}

\section{Introduction}
\label{sec:introduction}

For the problem of similarity search, we are given two texts and asked to find each sufficiently long substring of the first text that is within a certain Hamming distance of some substring of the second text.  Similarity search has many applications in bioinformatics --- e.g., ortholog detection, structure prediction or determining rearrangements --- and has been extensively studied~(see, e.g.,~\cite{SB05}).  Researchers used to first look for short substrings of the first text that occur unchanged in the second text, called seeds, then try to extend these short, exact matches in either direction to obtain longer, approximate matches.  This approach is called, naturally enough, ``seed and extend''.  The substrings' exact matches are found using either a hash table of the substrings with the right length, or an index structure such as a suffix array (SA).

Around the turn of the millenium, Burkhardt and K\"arkk\"ainen~\cite{BK03} and Ma, Tromp and Li~\cite{MTL02} independently proposed looking for short {\em subsequences} of the first text that have a certain shape and occur unchanged in the second text, and trying to extend those.  A binary string encoding the shape of a subsequence, with 1s indicating positions where the characters must match and 0s indicating positions where they need not, is called a spaced seed.  The total number of bits in the binary string is called the seed's length, and the number of 1s is called its weight.  The subsequences' exact matches are found using either a hash table of the subsequences with the right shape, or a kind of modified SA called a spaced suffix array~\cite{KWSHF11} (SSA).

Burkhardt and K\"arkk\"ainen, Ma et al.\ and subsequent authors have shown that using spaced seeds significantly improves the performance of seeding and extending.  Many papers have been written about how to design spaced seeds to minimize the number of errors (see, e.g.,~\cite{Bro08,EM13,IIKM11} and references therein), with the specifics depending on the model of sequence similarity and the acceptable numbers of false positives (for which the characters indicated by 1s all match but the substrings are not similar) and false negatives (for which those do not all match but the substrings are still similar) for the application in question.  Regardless of the particular application, however, researchers have consistently observed that the best results are obtained using more than one seed at a time.  A set of spaced seeds used in combination is called a multiple seed.

Multiple seeds are now a popular and powerful tool for similarity search, but they have a lingering flaw: we keep a hash table or SSA for each seed, and each instance of these data structures takes linear space.  For example, SHRiMP2's~\cite{DDLIB11} index for the human genome takes 16 GB {\em for each seed}.  In contrast, Bowtie 2's~\cite{LS12} compressed SA for that genome takes only 2.5 GB.  This is because a normal SA (which supports only substring matching) can be compressed such that the number of bits per character is only slightly greater than the empirical entropy of the text.  Unfortunately, the techniques for compressing normal SAs do not seem to apply directly to SSAs.

In this paper we show how to compress SSAs relative to normal SAs and still support fast random access to them.  Whereas the normal SA for a text lists the starting points of the suffixes of that text by those suffixes' lexicographic order, the SSA for a text and a spaced seed lists the starting points of the subsequences with the right shape by those subsequences' lexicographic order.  Intuitively, if the seed starts with many 1s, the SSA will be similar to the SA.  In Section~\ref{sec:theory} we formalize this intuition and prove a theoretical upper bound on the space needed to store an SSA when we already have the SA, in terms of the text's length, the alphabet's size, and the seed's length and weight.

In Section~\ref{sec:practice} we present experiments showing that our approach works even better in practice.  That is, even when we implement our data structures using simpler, theoretically sub-optimal components, we achieve better compression than our upper bound predicts.  In fact, in practice we can even successfully apply our approach in some cases when the assumptions underlying our upper bounds are violated.  However, we still want to improve our compression and random-access times for seeds with low weight-to-length ratios.

We recently learned that Peterlongo et al.~\cite{PPBS05} and Crochemore and Tischler~\cite{CT10} independently defined SSAs, under the names ``bi-factor arrays'' and ``gapped suffix arrays'', for the special case in which the spaced seed has the form \(1^a 0^b 1^c\).  Russo and Tischler~\cite{RT11} showed how to represent such an SSA in asymptotically succinct space such that we can support random access to it in time logarithmic in the length of the text.  We note, however, that the spaced seeds used for most applications do not have this form.  We also recently learned that Battaglia et al.~\cite{BCGP09} used an idea similar to that of spaced seeds in an algorithm for finding motifs with don't-care symbols.  It seems possible our results could be useful in reducing their algorithm's memory usage.

\section{Theory}
\label{sec:theory}

Suppose we want to store an SSA for a text \(T [0..n - 1]\) over an alphabet of size $\sigma$ and a spaced seed $S$ with length $\ell$ and weight $w$.  For \(i < n\), let \(T_i\) be the subsequence of \(T [i..n - 1]\) that contains \(T [j]\) if and only if \(i \leq j\) and \(S [j - i] = 1\).  Let \(T_i'\) be the subsequence of \(T [i..n - 1]\) that contains \(T [j]\) if and only if \(S [j - i] = 0\).  Let $\SSA$ be the permutation on \(\{0, \ldots, n - 1\}\) in which $i$ precedes $i'$ if either \(T_i \prec T_{i'}\), or \(T_i = T_{i'}\) and \(T [i..n - 1] \prec T [i'..n - 1]\).

For example, if \(T = \mathsf{abracadabra}\) and \(S = 101\) then
\[\begin{array}{rcl@{\hspace{5ex}}rcl@{\hspace{10ex}}rcl@{\hspace{5ex}}rcl}
T_0 & = & \mathsf{ar} & T_6 & = & \mathsf{db} & T_0' & = & \mathsf{b} & T_6' & = & \mathsf{a}\\
T_1 & = & \mathsf{ba} & T_7 & = & \mathsf{ar} & T_1' & = & \mathsf{r} & T_7' & = & \mathsf{b}\\
T_2 & = & \mathsf{rc} & T_8 & = & \mathsf{ba} & T_2' & = & \mathsf{a} & T_8' & = & \mathsf{r}\\
T_3 & = & \mathsf{aa} & T_9 & = & \mathsf{r} & T_3' & = & \mathsf{c} & T_9' & = & \mathsf{a}\\
T_4 & = & \mathsf{cd} & T_{10} & = & \mathsf{a} & T_4' & = & \mathsf{a} &&&\\
T_5 & = & \mathsf{aa} &&&& T_5' & = & \mathsf{d}\\
\end{array}\]
and so \(\SSA = [10, 3, 5, 7, 0, 8, 1, 4, 6, 9, 2]\), while \(\SA = [10, 7, 0, 3, 5, 8, 1, 4, 6, 9, 2]\).

If \(T_i \preceq T_{i'}\) and \(T_i' \preceq T_{i'}'\), then $i$ precedes $i'$ in both $\SSA$ and $\SA$.  In particular, if \(T_i = T_{i'}\) or \(T_i' = T_{i'}'\), then $i$ and $i'$ have the same relative order in $\SSA$ and $\SA$.  In our example, \(T_3 = T_5 = \mathsf{aa}\), so 3 precedes 5 in both $\SSA$ and $\SA$; \(T_2' = T_6' = T_9' = \mathsf{a}\), so 6 precedes 9 and 9 precedes 2 in both $\SSA$ and $\SA$.

If we partition $\SSA$ into subsequences such that $i$ and $i'$ are in the same subsequence if and only if \(T_i = T_{i'}\), then we can partition $\SA$ into the same subsequences.  Since there are at most \(\sigma^w + w\) distinct strings $T_i$, our partitions each consist of at most \(\sigma^w + w\) subsequences.  Similarly, if we partition based on $T_i'$ and $T_{i'}'$, then our partitions each consist of at most \(\sigma^{\ell - w} + \ell - w\) subsequences.

For our example, we can partition both $\SSA$ and $\SA$ into \([4, 6, 9, 2]\), for \(T_i' = \mathsf{a}\); \([7, 0]\), for \(T_i' = \mathsf{b}\); \([3]\), for \(T_i' = \mathsf{c}\); \([5]\), for \(T_i' = \mathsf{d}\); \([8, 1]\), for \(T_i' = \mathsf{r}\); and \([10]\), for \(T_i' = \epsilon\).  In this particular case, however, we could just as well partition both $\SSA$ and $\SA$ into only two common subsequences: e.g., \([10, 7, 0]\) and \([3, 5, 8, 1, 4, 6, 9, 2]\).

Consider the permutation \(\SA^{-1} \circ \SSA\), which maps elements' positions in $\SSA$ to their positions in $\SA$, and let $\rho$ be the minimum number of increasing subsequences into which \(\SA^{-1} \circ \SSA\) can be partitioned.  Since any subsequence common to $\SSA$ and $\SA$ corresponds to an increasing subsequence in \(\SA^{-1} \circ \SSA\), we have \(\rho \leq \min (\sigma^w + w, \sigma^{\ell - w} + \ell - w)\).  In our example, \(\SA^{-1} \circ \SSA = [0, 3, 4, 1, 2, 5, 6, 7, 8, 9, 10]\) and \(\rho = 2\).

Supowit~\cite{Sup85} gave a simple algorithm that partitions \(\SA^{-1} \circ \SSA\) into $\rho$ increasing subsequences in \(\Oh{n \lg \rho} \subseteq \Oh{n \min (w, \ell - w) \lg \sigma}\) time.  When applied to \(\SA^{-1} \circ \SSA\) in our example, Supowit's algorithm partitions it into \([0, 3, 4]\) and \([1, 2, 5, 6, 7, 8, 9, 10]\).

Barbay et al.~\cite{BCGNN??} showed how, given a partition of \(\SA^{-1} \circ \SSA\) into $\rho$ increasing subsequences, we can store it in \((2 + o (1)) n \lg \rho \leq (2 + o (1)) n \min (w, \ell - w) \lg \sigma\) bits and support random access to it in $\Oh{\lg \lg \rho}$ time.  Combining their ideas with later work by Belazzougui and Navarro~\cite{BN??}, we can keep the same space bound and improve the time bound to $\Oh{1}$.

To do this, for \(i \leq \rho\), we replace each element in the $i$th subsequence in \(\SA^{-1} \circ \SSA\) by a character $a_i$, and store the resulting string $R$ such that we can support random access to it and partial rank queries on it.  We then permute $R$ according to \(\SA^{-1} \circ \SSA\) and store the resulting string $R'$ such that we can support fast select queries on it.  In our example, \(R = a_1 a_2 a_2 a_1 a_1 a_2 a_2 a_2 a_2 a_2 a_2\) and \(R' = a_1 a_1 a_1 a_2 a_2 a_2 a_2 a_2 a_2 a_2 a_2\).

The partial rank query \(R.\prank (i)\) returns the number of copies of \(R [i]\) in \(R [0..i]\), and the select query \(R'.\select_a (i)\) returns the position of the $i$th copy of $a$ in $R'$.  Barbay et al.\ noted that, for \(i < n\),
\[(\SA^{-1} \circ \SSA) [i] = R'.\select_{R [i]} (R.\prank (i))\ .\]
Belazzougui and Navarro showed how we can store $R$ in \((1 + o (1)) n \lg \rho\) bits and support random access to it and partial rank queries on it in $\Oh{1}$ time, and store $R'$ in \((1 + o (1)) n \lg \rho\) bits and support select queries on it in $\Oh{1}$ time.

In summary, we can store \(\SA^{-1} \circ \SSA\) in \((2 + o (1)) n \min (w, \ell - w) \lg \sigma\) bits such that we can support random access to it in $\Oh{1}$ time.  We will give a longer explanation in the full version of this paper.  Since \(\SSA = \SA \circ (\SA^{-1} \circ \SSA)\), this gives us the following result:

\begin{theorem}
\label{thm:compressed}
Let \(T [0..n - 1]\) be a text over an alphabet of size $\sigma$ and let $S$ be a spaced seed with length $\ell$ and weight $w$.  If we have already stored the suffix array $\SA$ for $T$ such that we can support random access to $\SA$ in time $t_\SA$, then we can store a spaced suffix array $\SSA$ for $T$ and $S$ in \((2 + o (1)) n \min (w, \ell - w) \lg \sigma\) bits such that we can support random access to $\SSA$ in \(t_\SA + \Oh{1}\) time.
\end{theorem}

\section{Practice}
\label{sec:practice}

Theorem~\ref{thm:compressed} says we can store SSAs for the human Y-chromosome {\tt chrY.fa} in FASTA format (about 60 million characters over an alphabet of size 5) and SHRiMP2's three default spaced seeds --- i.e., 11110111101111, 1111011100100001111 and 1111000011001101111 --- in about 560 MB, in addition to the SA, whereas storing the SSAs na\"ively would take about 720 MB.  Storing the SSAs packed such that each entry takes \(\lceil \lg 60\,000\,000 \rceil = 26\) bits would reduce this to about 580 MB.

To test our approach, we built the SSAs as described in Section~\ref{sec:theory}; computed \(\SA^{-1} \circ SSA\), $R$ and $R'$ for each SSA; and stored each copy of $R$ or $R'$ as a wavelet tree.  We chose wavelet trees because they are simple to use and often more practical than the theoretically smaller and faster data structures mentioned in Section~\ref{sec:theory}.  We ran all our tests described in this section on a computer with a quad-core Intel Xeon CPU with 32 GB of RAM, running Ubuntu 12.04.  We used a wavelet-tree implementation from {\tt https://github.com/fclaude/libcds} and compiled it with GNU {\tt g++} version 4.4.3 with optimization flag {\tt -O3}.

The uncompressed SA took 226 MB, and the six wavelet trees took a total of 215 MB and performed 10\,000 random accesses each in 7.67 microseconds per access.  That is, we compressed the SSAs into about 60\% of the space it would take to store them na\"ively and, although our accesses were much slower than direct memory accesses, they were fast compared to disk accesses.  Thus, our approach seems likely to be useful when a set of SSAs is slightly larger than the memory and fits only when compressed.

Using the same test setup, we then compressed SSAs for the ten spaced seeds BFAST~\cite[Table S3]{HMN09} uses for 36-base-pair Illumina reads, which all have weight 18:
{\small
\begin{enumerate}
\item 111111111111111111
\item 11110100110111101010101111
\item 11111111111111001111
\item 1111011101100101001111111
\item 11110111000101010000010101110111
\item 1011001101011110100110010010111
\item 1110110010100001000101100111001111
\item 1111011111111111111
\item 11011111100010110111101101
\item 111010001110001110100011011111\,.
\end{enumerate}}
Since the first seed consists only of 1s, the SSA we would build for it is the same as the SA.  The uncompressed SA again took 226 MB and the 18 wavelet trees for the other nine seeds took a total of 649 MB --- so instead of 2.26 GB, we used 875 MB (about 39\%) for all ten seeds --- and together performed 10\,000 random accesses to each of the ten SSAs in about 7 microseconds per access.  The left side of the top half of Figure~\ref{fig:bfast} shows how many bits per character (bpc) of the text each SSA took, and the average time per access to each SSA.

We also compressed the SSAs for the ten spaced seeds BFAST uses for 50-base-pair Illumina reads, which all have weight 22:
{\small
\begin{enumerate}
\item 1111111111111111111111
\item 1111101110111010100101011011111
\item 1011110101101001011000011010001111111
\item 10111001101001100100111101010001011111
\item 11111011011101111011111111
\item 111111100101001000101111101110111
\item 11110101110010100010101101010111111
\item 111101101011011001100000101101001011101
\item 1111011010001000110101100101100110100111
\item 1111010010110110101110010110111011\,.
\end{enumerate}}
Again, the first seed consists only of 1s.  This time, the 18 wavelet trees for the other nine seeds took a total of 712 MB; each access took about 8 microseconds.  The left side of the bottom half of Figure~\ref{fig:bfast} shows how many bit per character of the text each SSA took, and the average access time per access to each SSA.

If we have a permutation $\pi_1$ on \(\{0, \ldots, n - 1\}\) stored and $\pi_2$ is any other permutation on \(\{0, \ldots, n - 1\}\), then we can store $\pi_2$ relative to $\pi_1$ using the ideas from Section~\ref{sec:theory}.  For example, we can store SSAs relative to other SSAs.  Suppose we consider the size of each SSA (except the SA) when compressed relative to each other SSA (including the SA), build a minimum spanning tree rooted at the SA, and compress each SSA relative to its parent in the tree.  This can reduce our space usage at the cost of increasing the random-access time, as shown for the BFAST seeds on the right side of Figure~\ref{fig:bfast}.

\begin{figure}[t]
\begin{centering}
\resizebox{55ex}{!}
{\begin{tabular}{c@{\hspace{15ex}}c}
\begin{tabular}{r@{\hspace{3ex}}r@{\hspace{3ex}}r}
& space & time\\
seed & (bpc) & ($\mu$s)\\[0.5ex]
\hline\\[-1.5ex]
1 & 32.00 & 0 \\
2 & 11.29 & 9 \\
3 & 4.41 & 4 \\
4 & 9.75 & 8 \\
5 & 11.54 & 9 \\
6 & 13.77 & 11 \\
7 & 13.14 & 10 \\
8 & 3.85 & 3 \\
9 & 10.10 & 7 \\
10 & 13.91 & 11 \\
\end{tabular} &
\begin{tabular}{r@{\hspace{3ex}}c@{\hspace{3ex}}r@{\hspace{3ex}}r}
&& space & time\\
seed & reference & (bpc) & ($\mu$s)\\[0.5ex]
\hline\\[-1.5ex]
1 &  -  &  32.00 &  0\\
2 & 8 & 9.71  & 11\\
3 & 1 & 4.41  & 4\\
4 & 8 & 9.22  & 10\\
5 & 4 & 9.23  & 19\\
6 & 8 & 12.27 & 14\\
7 & 3 & 12.58 & 14\\
8 & 1 & 3.85  & 3\\
9 & 1 & 10.10 & 7\\
10 & 7 & 12.59 & 26\\
\end{tabular}\\\rule{0ex}{5ex}&\\
\begin{tabular}{r@{\hspace{3ex}}r@{\hspace{3ex}}r}
& space & time\\
seed & (bpc) & ($\mu$s)\\[0.5ex]
\hline\\[-1.5ex]
1 &32.00&0\\
2 &9.03&8\\
3 &12.30&10\\
4 &13.86&11\\
5 &8.13&7\\
6 &10.80&9\\
7 &11.14&8\\
8 &11.09&8\\
9 &11.77&9\\
10 &12.54&10
\end{tabular} &
\begin{tabular}{r@{\hspace{3ex}}c@{\hspace{3ex}}r@{\hspace{3ex}}r}
&& space & time\\
seed & reference & (bpc) & ($\mu$s)\\[0.5ex]
\hline\\[-1.5ex]
1&- & 32.00& 0\\
2&1 & 9.03& 6\\
3&1 & 12.30& 10\\
4&2 & 12.59& 18\\
5&1 & 8.13& 6\\
6&1 & 10.80& 8\\
7&1 & 11.14& 8\\
8&1 & 11.09& 9\\
9&8 & 11.34& 18\\
10&8 & 8.94& 17
\end{tabular}
\end{tabular}}
\caption{The space usage of the SSAs of the spaced seeds BFAST uses for Illumina reads, in bits per character of the text, and the average time for a random access.  On top, the seeds are for 36-base-pair reads; on the bottom, the seeds are for 50-base-pair reads.  On the left, all the SSAs are compressed relative to the SA; on the right, some of the SSAs are compressed relative to other SSAs.}
\label{fig:bfast}
\end{centering}
\end{figure}

There are other circumstances in which we can ignore SSAs' semantics and consider them only as permutations.  For example, spaced seeds can be generalized to subset seeds~\cite{KNR06}, such as ternary strings in which 1s indicate positions where the characters must match, 0s indicate positions where they need not, and Ts indicate positions where characters must fall within the same equivalence class (such as the pyrimidines {\tt C} and {\tt T} and the purines {\tt A} and {\tt G}).  It is not difficult to generalize Theorem~\ref{thm:compressed} to subset seeds --- we will do so in the full version of this paper --- but it is also not necessary to obtain practical results.  The {\tt Iedera} tool (available at {\tt http://bioinfo.lifl.fr/yass/iedera.php}) generates good subset seeds.

A more challenging change is from fixed-length seeds to repetitive seeds~\cite{KWSHF11}.  A repetitive seed is a string in whose repetition the digits indicate which characters must match and how.  For example, with respect to the repetitive spaced seed 10110, {\tt ATCGATCGGT} matches {\tt ACCGTTGGGA} but not {\tt ACCGTTGAGA}.  Repetitive seeds are useful when looking for approximate matches of substrings that have been extended until they become sufficiently infrequent.  It is not clear how or if we can extend Theorem~\ref{thm:compressed} to repetitive seeds.  Nevertheless, the {\tt LAST} tool (available at {\tt http://last.cbrc.jp}) generates SSAs for repetitive spaced or subset seeds, which we can still try to compress in practice; see also~\cite{HKF08,OS11}.

Our current goal is to achieve reasonable compression and access times for a set of repetitive subset seeds that we received from Martin Frith, which have average length 19.85 and average weight about 10.44, counting ``same equivalence class'' digits as 0.5.  Unfortunately, at the moment we use nearly 24 bits per entry in the corresponding SSAs (including the overhead for the uncompressed SA), which is only marginally better than the 26 bits we would use with simple packing.  Meanwhile, random accesses take about 12 microseconds on average, which is significantly slower than access to a packed array.  On the other hand, these seeds have an unusually low average weight-to-length ratio.  We used {\tt Iedera} and {\tt LAST} to generate SSAs for a set of eight repetitive subset seeds, with average length 17.875 and average weight 12.  For these, we used only 20.15 bits per entry, with random accesses taking about 10 microseconds on average.

\section{New Directions}
\label{sec:new}

Barbay et al.'s result holds even when we are given a partition of a permutation into $\rho$ increasing or decreasing subsequences, and some authors~\cite{ACDD+11,CM13} have found that using both increasing and decreasing subsequences often improves compression in practice.  Computing a partition into the minimum number of such subsequences is NP-hard, however, and we see no reason why \(\SA^{-1} \circ \SSA\) should contain long decreasing subsequences.  Therefore, in this paper we considered only increasing subsequences and did not discuss other papers on compressing permutations (see~\cite{BN13} and references therein).

Nevertheless, while writing this paper we noticed a way to generalize slightly Barbay et al.'s result.  Suppose we partition a permutation \(\pi [0..n - 1]\) into subsequences \(\tau_0, \ldots, \tau_{\rho - 1}\).  Let \(R [0..n - 1]\) be the bitvector in which \(R [i] = j\) if \(\pi [i]\) is in $\tau_j$.  Let \(R' [0..n - 1]\) be the bitvector in which \(R' [i] = j\) if $i$ is in $\tau_j$.  Let \(\pi_0 [0..|\tau_0| - 1], \ldots, \pi_{\rho - 1} [0..|\tau_{\rho - 1}| - 1]\) be the permutations in which \(\pi_i [j]\) is the number of elements strictly smaller than \(\tau_i [j]\) in $\tau_i$.  Then
\[\pi [i] = R'.\select_{R [i]} \left( \pi_{R [i]} \left[ \rule{0ex}{2ex} R.\prank (i) - 1 \right] + 1 \right)\,.\]
Therefore, if we can store \(\pi_0, \ldots, \pi_{\rho - 1}\) in small space, then we can store $\pi$ in small space.  Notice also that, conversely,
\[\pi_{j} [i] = R'.\prank \left( \pi \left[ \rule{0ex}{2ex} R.\select_j (i + 1) \right] \right) - 1\,;\]
this will be useful later.

For example, if
\begin{eqnarray*}
\pi [0..19] & = & [16, 17, 12, 10, 6, 1, 9, 15, 18, 4, 14, 13, 5, 11, 19, 2, 8, 7, 0, 3]\\
\tau_0 [0..8] & = & [16, 12, 6, 1, 15, 13, 5, 2, 0]\\
\tau_1 [0..10] & = & [17, 10, 9, 18, 4, 14, 11, 19, 8, 7, 3]
\end{eqnarray*}
then
\begin{eqnarray*}
R [0..19] & = & 01010010111001101101\\
R' [0..19] & = & 00011001111100100111\\
\pi_0 [0..8] & = & [8, 5, 4, 1, 7, 6, 3, 2, 0]\\
\pi_1 [0..10] & = & [8, 5, 4, 9, 1, 7, 6, 10, 3, 2, 0]
\end{eqnarray*}
and thus, say,
\begin{eqnarray*}
\pi [6]
& = & R'.\select_{R [i]} \left( \pi_{R [i]} \left[ \rule{0ex}{2ex} R.\prank (i) - 1 \right] + 1 \right)\\
& = & R'.\select_1 (\pi_1 [2] + 1)\\
& = & R'.\select_1 (5)\\
& = & 9\,.
\end{eqnarray*}
Therefore, since the elements in $\pi_0$ and $\pi_1$ appear in lexicographic order by name, we can store $\pi$ in small space.

This may seem at first like rather a pointless generalization, but consider the case in which we want to store a permutation $\pi$ that is similar to a permutation $\widehat{\pi}$ that we already have stored.  By similar, we mean here that we can find a long subsequences in $\pi$ and $\widehat{\pi}$ such that the elements in those subsequences have the same relative order.  In this case, we can use use $\widehat{\pi}$ to compress $\pi$.

For example, if
\begin{eqnarray*}
\pi [0..12] & = & [12, 11, 6, 3, 8, 0, 5, 7, 4, 9, 1, 10, 2]\\
\widehat{\pi} [0..11] & = & [11, 10, 7, 0, 3, 5, 8, 1, 4, 6, 9, 2]
\end{eqnarray*}
then the elements in their subsequences \([12, 11, 8, 0, 9, 1, 10, 2]\) and \([11, 10, 7, 0, 8, 1, 9, 2]\) have the same relative order, \([7, 6, 3, 0, 4, 1, 5, 2]\).  Let
\begin{eqnarray*}
\tau_0 [0..7] & = & [12, 11, 8, 0, 9, 1, 10, 2]\\
\tau_1 [0..4] & = & [6, 3, 5, 7, 4]\\
\widehat{\tau}_0 [0..7] & = & [11, 10, 7, 0, 8, 1, 9, 2]\\
\widehat{\tau}_1 [0..3] & = & [3, 5, 4, 6]
\end{eqnarray*}
so
\begin{eqnarray*}
R [0..12] & = & 0011001110000\\
R' [0..12] & = & 0001111100000\\
\widehat{R} [0..11] & = & 000011001100\\
\widehat{R}' [0..11] & = & 000111100000\\
\pi_0 [0..7] = \widehat{\pi}_0 [0..7] & = & [7, 6, 3, 0, 4, 1, 5, 2]\\
\pi_1 [0..4] & = & [3, 0, 2, 4, 1]\\
\widehat{\pi}_1 & = & [0, 2, 1, 3]\,,
\end{eqnarray*}
where we use accents to distinguish the data related to $\widehat{\pi}$ from those related to $\pi$.

Because we already have $\widehat{\pi}$ stored, if we store $\widehat{R}$ and $\widehat{R}'$, then for \(0 \leq i \leq 7\) we can compute in $\Oh{1}$ time
\[\pi_0 [i]
= \widehat{\pi}_0 [i]
= \widehat{R}'.\prank \left( \widehat{\pi} \left[ \rule{0ex}{2ex} \widehat{R}.\select_0 (i + 1) \right] \right) - 1\,.\]
Therefore, if we also store $R$, $R'$ and $\pi_1$, then we can support access to $\pi$ in $\Oh{1}$ time.  That is, we can store \(\pi [0..12]\) as \(\tau_1 [0..4]\) and four bitvectors.

We chose \(\pi = [12, 11, 6, 3, 8, 0, 5, 7, 4, 9, 1, 10, 2]\) and \(\widehat{\pi} = [11, 10, 7, 0, 3, 5, 8, 1, 4, 6, 9, 2]\) because they are the SAs of {\sf abrabbababra\$} and {\sf abracadabra\$}.  Thus, our example illustrates how we can usually store the SA of a slightly modified version of a document in small space when we already have the SA of the original: we look for a long subsequence of the new SA in which the elements have the same order as those in a subsequence of the original SA, and store the complement of that subsequence and the appropriate four bitvectors.  Finding the longest subsequence to use is NP-hard, by a reduction from permutation pattern matching~\cite{BBL98}, but L\'eonard, Mouchard and Salson~\cite{LMS12} showed how we can find a reasonably long one in the average case and in practice.

As a very preliminary test, we built the SA for the Wikipedia page for ``Suffix array'' \linebreak ({\tt http://en.wikipedia.org/wiki/Suffix$\_$array}, downloaded on February 13th, 2014) and the SA for the same page with the order of the bullet points swapped in the paragraph
\begin{quotation}
\noindent ``Suffix arrays are closely related to suffix trees:
\begin{itemize}
\item Suffix arrays can be constructed by performing a depth-first traversal of a suffix tree. The suffix array corresponds to the leaf-labels given in the order in which these are visited during the traversal, if edges are visited in the lexicographical order of their first character.
\item A suffix tree can be constructed in linear time by using a combination of suffix and LCP array. For a description of the algorithm, see the corresponding section in the LCP array article.''
\end{itemize}
\end{quotation}
Using the approach described above, we were able to store the SA of the modified page in about 0.4 bits per character, on top of the SA of the original page.  In comparison, compressing the modified page with {\tt p7zip} (available at {\tt https://packages.debian.org/sid/p7zip-full}) used about 1.85 bits per character.

We hope our approach will prove useful for storing and indexing, e.g., many separate human genomes.  (We note that indexing the concatenation of many human genomes has been studied previously; see, e.g.,~\cite{MNSV10}.)  Since human genomes are similar to each other, we should be able to support fast access to a (possibly sampled) SA for each genome while using reasonable total space.  Moreover, most of the genomes' Burrows-Wheeler Transforms~\cite{BW94} (BWTs) should share a long common subsequence, so we should be able to store also rank and select data structures for them in reasonable total space: for each BWT, we store a bitvector indicating which characters are part of the common subsequence, and a rank and select data structure for those characters not in the common subsequence; we then need store only a single rank and select data structure for the common subsequence.  Therefore, we should be able to store an FM-index~\cite{FM05} for each genome using reasonable total space.  Investigating this possibility is beyond the scope of this paper, however, so we leave it as future work.

\section*{Acknowledgments}

Many thanks to Francisco Claude, Maxime Crochemore, Matei David, Martin Frith, Costas Iliopoulos, Juha K\"arkk\"ainen, Gregory Kucherov, Bin Ma, Ian Munro, Taku Onodera, Gonzalo Navarro, Luis Russo, German Tischler and the anonymous reviewers.

\bibliographystyle{plain}
\bibliography{cssa}

\begin{thebibliography}{10}

\bibitem{ACDD+11}
D.~Arroyuelo, F.~Claude, R.~Dorrigiv, S.~Durocher, M.~He, A.~{L\'opez-Ortiz},
  J.~I. Munro, P.~K. Nicholson, A.~Salinger, and M.~Skala.
\newblock Untangled monotonic chains and adaptive range search.
\newblock {\em Theoretical Computer Science}, 432:4200--4211, 2011.

\bibitem{BCGNN??}
J.~Barbay, F.~Claude, T.~Gagie, G.~Navarro, and Y.~Nekrich.
\newblock Efficient fully-compressed sequence representations.
\newblock {\em Algorithmica}.
\newblock To appear.

\bibitem{BN13}
J.~Barbay and G.~Navarro.
\newblock On compressing permutations and adaptive sorting.
\newblock {\em Theoretical Computer Science}, 513:109--123, 2013.

\bibitem{BCGP09}
G.~Battaglia, D.~Cangelosi, R.~Grossi, and N.~Pisanti.
\newblock Masking patterns in sequences: A new class of motif discovery with
  don't cares.
\newblock {\em Theoretical Computer Science}, 410:4327--4340, 2009.

\bibitem{BN??}
D.~Belazzougui and G.~Navarro.
\newblock Alphabet-independent compressed text indexing.
\newblock {\em ACM Transactions on Algorithms}.
\newblock To appear.

\bibitem{BBL98}
P.~Bose, J.~F. Buss, and A.~Lubiw.
\newblock Pattern matching for permutations.
\newblock {\em Information Processing Letters}, 65:277--283, 1998.

\bibitem{Bro08}
D.~G. Brown.
\newblock A survey of seeding for sequence alignment.
\newblock In I.~M{\v{a}}ndoiu and A.~Zelikovsky, editors, {\em Bioinformatics
  Algorithms: Techniques and Applications}, pages 126--152. Wiley-Interscience,
  2008.

\bibitem{BK03}
S.~Burkhardt and J.~K{\"a}rkk{\"a}inen.
\newblock Better filtering with gapped q-grams.
\newblock {\em Fundamenta Informaticae}, 56:51--70, 2003.

\bibitem{BW94}
M.~Burrows and D.~J. Wheeler.
\newblock A block sorting lossless data compression algorithm.
\newblock Technical Report 124, Digital Equipment Corporation, 1994.

\bibitem{CM13}
F.~Claude and J.~I. Munro.
\newblock Adaptive data structures for permutations and binary relations.
\newblock In {\em Proceedings of the 20th Symposium on String Processing and
  Information Retrieval (SPIRE)}, pages 64--71, 2013.

\bibitem{CT10}
M.~Crochemore and G.~Tischler.
\newblock The gapped suffix array: A new index structure for fast approximate
  matching.
\newblock In {\em Proceedings of the 17th Symposium on String Processing and
  Information Retrieval (SPIRE)}, pages 359--364, 2010.

\bibitem{DDLIB11}
M.~David, M.~Dzamba, D.~Lister, L.~Ilie, and M.~Brudno.
\newblock {SHRiMP2}: Sensitive yet practical short read mapping.
\newblock {\em Bioinformatics}, 27:1011--1012, 2011.

\bibitem{EM13}
L.~Egidi and G.~Manzini.
\newblock Better spaced seeds using quadratic residues.
\newblock {\em Journal of Compututer and System Sciences}, 79:1144--1155, 2013.

\bibitem{FM05}
P.~Ferragina and G.~Manzini.
\newblock Indexing compressed text.
\newblock {\em Journal of the ACM}, 52:552--581, 2005.

\bibitem{HMN09}
N.~Homer, B.~Merriman, and S.~F. Nelson.
\newblock {BFAST}: An alignment tool for large scale genome resequencing.
\newblock {\em PLOS One}, 4:e7767, 2009.

\bibitem{HKF08}
P.~Horton, S.~M. Kie{\l}basa, and M.~C. Frith.
\newblock {DisLex}: A tranformation for discontiguous suffix array
  construction.
\newblock In {\em Proceedings of the Workshop on Knowledge, Language, and
  Learning in Bioinformatics (KLLBI)}, pages 1--11, 2008.

\bibitem{IIKM11}
L.~Ilie, S.~Ilie, S.~Khoshraftar, and A.~{Mansouri Bigvand}.
\newblock Seeds for effective oligonucleotide design.
\newblock {\em BMC Genomics}, 12:280, 2011.

\bibitem{KWSHF11}
S.~M. Kie{\l}basa, R.~Wan, K.~Sato, P.~Horton, and M.~C. Frith.
\newblock Adaptive seeds tame genomic sequence comparison.
\newblock {\em Genome Research}, 21:487--493, 2011.

\bibitem{KNR06}
G.~Kucherov, L.~No{\'e}, and M.~A. Roytberg.
\newblock A unifying framework for seed sensitivity and its application to
  subset seeds.
\newblock {\em Journal of Bioinformatics and Computational Biology},
  4:553--570, 2006.

\bibitem{LS12}
B.~Langmeand and S.~L. Salzberg.
\newblock Fast gapped-read alignment with {Bowtie} 2.
\newblock {\em Nature Methods}, 9:357--359, 2012.

\bibitem{LMS12}
M.~L{\'e}onard, L.~Mouchard, and M.~Salson.
\newblock On the number of elements to reorder when updating a suffix array.
\newblock {\em Journal of Discrete Algorithms}, 11:87--99, 2012.

\bibitem{MTL02}
B.~Ma, J.~Tromp, and M.~Li.
\newblock {PatternHunter}: faster and more sensitive homology search.
\newblock {\em Bioinformatics}, 18:440--445, 2002.

\bibitem{MNSV10}
V.~M{\"a}kinen, G.~Navarro, J.~Sir{\'e}n, and N.~V{\"a}lim{\"a}ki.
\newblock Storage and retrieval of highly repetitive sequence collections.
\newblock {\em Journal of Computational Biology}, 17:281--308, 2010.

\bibitem{OS11}
T.~Onodera and T.~Shibuya.
\newblock An index structure for spaced seed search.
\newblock In {\em Proceedings of the 22nd International Symposium on Algorithms
  and Computation (ISAAC)}, pages 764--772, 2011.

\bibitem{PPBS05}
P.~Peterlongo, N.~Pisanti, F.~Boyer, and M.-F. Sagot.
\newblock Lossless filter for finding long multiple approximate repetitions
  using a new data structure, the bi-factor array.
\newblock In {\em Proceedings of the 12th Symposium on String Processing and
  Information Retrieval (SPIRE)}, pages 179--190, 2005.

\bibitem{RT11}
L.~M.~S. Russo and G.~Tischler.
\newblock Succinct gapped suffix arrays.
\newblock In {\em Proceedings of the 18th Symposium on String Processing and
  Information Retrieval (SPIRE)}, pages 290--294, 2011.

\bibitem{SB05}
Y.~Sun and J.~Buhler.
\newblock Designing multiple simultaneous seeds for {DNA} similarity search.
\newblock {\em Journal of Computational Biology}, 12:847--861, 2005.

\bibitem{Sup85}
K.~J. Supowit.
\newblock Decomposing a set of points into chains, with applications to
  permutation and circle graphs.
\newblock {\em Information Processing Letters}, 21:249--252, 1985.

\end{thebibliography}

\end{document}